\documentclass[a4paper,fleqn,usenatbib]{aa}
\usepackage{txfonts}

\usepackage{graphicx}	
\usepackage{subfigure}  
\usepackage{amsmath}	
\usepackage{amssymb}	
\usepackage{bm}
\usepackage{url}
\usepackage{upgreek}
\usepackage{xspace}
\usepackage{xcolor}
\usepackage{hyperref}
\hypersetup{
    pdfborder={0 0 0 0},
    colorlinks=true,
    linkcolor=blue,
    urlcolor=blue,
    citecolor=blue
}

\newcommand{\rficlean}{\texttt{\mbox{RFIClean}}\xspace}
\newcommand{\jpsr}{J0437$-$4715\xspace}
\newcommand{\tdft}{2D DFT\xspace}
\newcommand{\rfrb}{FRB 20180916B\xspace}
\newcommand{\mean}{\ensuremath{\upmu}\xspace}
\newcommand{\sig}{\ensuremath{\sigma}\xspace}
\newcommand{\nt}{N\ensuremath{_t}\xspace}
\newcommand{\nf}{N\ensuremath{_f}\xspace}

\begin{document}


\title{Fourier domain excision of periodic radio frequency interference}

\author{
Yogesh Maan\inst{1}\fnmsep\thanks{E-mail: maan@astron.nl (YM)} \and
Joeri van Leeuwen\inst{1,2} \and
Dany Vohl\inst{1}
}

  \institute{ASTRON, the Netherlands Institute for Radio
  Astronomy, Postbus 2, 7990 AA, Dwingeloo, The Netherlands
  \and
  Anton Pannekoek Institute for Astronomy, University of
  Amsterdam, Science Park 904, 1098 XH, Amsterdam, The Netherlands
}

\date{Accepted XXX. Received YYY; in original form ZZZ}

\abstract{
The discovery and study of pulsars and Fast Radio Bursts (FRBs) in time-domain radio data is often hampered by radio frequency interference (RFI).
Some of this terrestrial RFI is impulsive and bright, and relatively easy to identify and remove.
Other anthropogenic signals, however, are weaker yet periodic, and their persistence can drown out astrophysical signals.
Here we show that Fourier-domain excision of periodic RFI is an effective and powerful step in detecting weak cosmic signals. 
We find that applying the method significantly increases the signal-to-noise ratio of transient as well as periodic pulsar signals.    
In live studies, we detected single pulses from pulsars and FRBs that would otherwise have remained buried in background noise. 
We show the method has no negative effects on pulsar pulse shape and that it enhances timing campaigns. 
We demonstrate the method on real-life data from number of large radio telescopes, and 
conclude that Fourier-domain RFI excision increases the effective sensitivity to astrophysical sources
by a significant fraction which can be even larger than an order of magnitude in case of strong RFI.
An accelerated implementation of the method runs on standard time-domain radio data formats and is publicly available.
}

\maketitle

\section{Introduction}
As both the aperture area and the bandwidth of radio telescopes increase,
the sensitivity
with which the radio sky can be probed continues to increase. However,
the potentially achievable sensitivity is often severely limited by
the ever increasing radio frequency interference (RFI). Preventive
methods, such as using remote, RFI-free locations for constructing the
new telescopes, identifying and isolating the sources of RFI
\citep[][]{2016RSEnv.180...76A}, and avoiding the known RFI-prone parts
of the spectrum \citep[e.g.,][]{Maan13}, remain crucial for obtaining
useful data from radio telescopes. However, even after exploiting these
and other preventive methods, data obtained from radio telescopes are
often significantly contaminated by RFI. Much of the radio spectrum used
by new broad-band receivers is outside the protected spectrum. The ever
more quickly growing number of radio-emitting satellites affect telescopes
even in radio-quiet and remote zones. The efficient mitigation of such RFI
remains a challenging process. 
\par
The RFI mitigation methods are broadly divided in two categories:
pre-detection and post-detection strategies. The pre-detection techniques
often use reference signals or relatively simple statistical criteria to
identify and mitigate RFI signals in real-time \citep[e.g.,][]{Baan04,Buch16}.
These techniques are often more suitable for highly impulsive or narrow-band
RFI. The hard requirement of real-time baseband processing might also pose
implementation challenges, especially for wide-band systems and existing
backends. The post-detection techniques mostly involve off-line processing,
though some rare but useful examples of real-time excision also exist
\citep[e.g.][]{sclocco_real-time_2020}.
These strategies often use thorough statistical techniques or exploit the
expected signatures, either of the signal-of-interest or that of
the RFI signal, to identify and mitigate RFI. There is no one unique technique
that can be used to mitigate all kinds of RFI, given the ever growing number
of RFI sources and the huge diversity in the effects in which they manifest
themselves in the radio astronomical data. On one side,
 mitigation techniques include identifying outliers using the first and
second order moments or other simple Gaussian statistics
\citep[e.g.][]{Baddi11}; or using higher order statistics
\citep{Dwyer83} such as the spectral kurtosis \citep{Nita07,NG10a}.
On the other side, methods focus on specific types of astronomical
\citep[e.g.][]{EKL09,Hogden12} or RFI signals where otherwise a
robust statistics approach might fail
\citep[e.g.][]{Deshpande05,Lazarus2020ascl.soft03008L}.
\par
Many robust statistical approaches have been developed to excise
primarily the RFI instances with spiky or burst-like characteristics.
These techniques are not readily applicable  when the data are
heavily affected by \emph{periodic} RFI. One well known source of
periodic interference is the overhead 50/60\,Hz power-lines which
could produce radio noise by partial discharge at insulators and the
sharp points of conductors. Moreover, most of the household and
commercial electrical appliances also operate on the 50/60\,Hz AC,
and if not properly designed or shielded, these might generate
interference with 50/60\,Hz periodicity. Several other sources,
such as military, aviation, airport and even ionosphere-research
radars could also cause periodic interference, though the periods of
such interference are hard to know in advance. Electric fences also
cause interference with nearly 1\,second periodicity, but the actual
period could differ depending on the design.
Windmills also cause periodic interference, though the periods could be
as short as a few microseconds. The periods of interference from these, and
many unknown sources could also vary significantly with time.
%
\begin{figure*}
\centering
\includegraphics[width=0.27\textwidth,angle=-90]{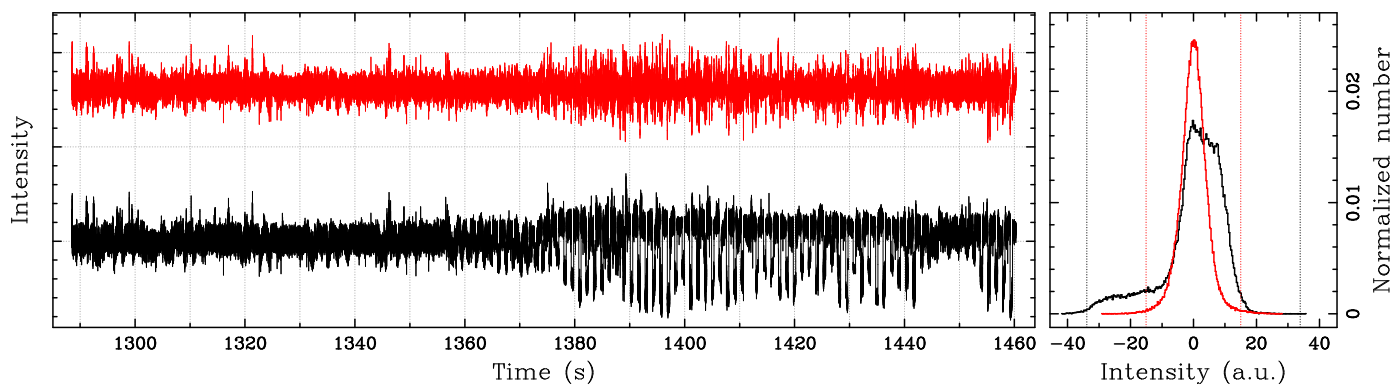}
\caption{The left panel shows the same parts of two intensity time-sequences.
The black colored one is the original sequence recorded
at a radio telescope, the second half of which is evidently contaminated
by periodic RFI. The red colored one shows the resultant sequence when 
the strong frequency components in the Fourier transform of the black
colored one are zapped. Both the sequences are detrended for any slow
variations before plotting here. The right panel shows the normalized
histograms for the two sequences shown on the left and the
$\pm$3.5$\sigma$ bounds around the mean level marked as vertical lines,
in the corresponding colors.}
\label{fig-demo}
\end{figure*}
%
\par
Contamination by periodic interference changes the apparent nature of
the underlying statistical distribution, making it harder to identify
RFI using conventional techniques.
Moreover, in case of such contaminations, the excision techniques
which identify the strong periodic interference, such as \texttt{rfifind}
from the pulsar search and analysis software \textsl{PRESTO} \citep{Ransom02},
could potentially result in masking a large fraction of data. In case of
periodic RFI, it is possible to largely remove the contribution of the
periodic interference from data, and hence avoid masking a large fraction
as RFI-contaminated.
In this work, we describe a technique
and present software tools to excise periodic interference in the
Fourier domain, restore the underlying statistics, and then use the
conventional methods for further RFI mitigation. We demonstrate the
efficacy of this approach using real-life observations of primarily two
types of astronomical sources, radio pulsars and Fast Radio Bursts (FRBs).
Pulsars are fast rotating neutron stars \citep{Hewish68}, and the radio
emission from them is observed in the form of a highly periodic sequence
of regular pulses. FRBs are milliseconds-wide highly
luminous radio transient events, most-likely of extra-galactic origin
\citep{Lorimer07,Thornton13}. While the presented technique and software
are readily applicable to pulsar and FRB data, the underlying approach
can be applied to time-domain radio data on any other astronomical source.
\par
In what follows, we elaborate on the impact of periodic RFI and its
mitigation using the one-dimensional Fourier transform in Section~2. In
Section~3, we present
details of a software tool \rficlean that we have developed to
excise periodic and other RFI from time-domain data in the widely used filterbank
format. We then briefly discuss mitigating wide-band, fainter periodic
interference using the two-dimensional Fourier transform and other future
work in section 4, followed by conclusions in the last section.
\section{Periodic RFI: Impact and mitigation}
For a time-sequence of radio intensity or power data, $f(\rm t)$, the
impulsive RFI-contaminated samples generally lie significantly outside
the expected distribution of the non-contaminated samples. With averaging
taking place at several stages in the data-recording systems, the
distribution of the final time-sequence is often very close to a
Gaussian function.
However, the presence of strong periodic RFI can significantly
change the underlying Gaussian statistics.
\par
Most RFI excision techniques heavily use the first two moments of the
underlying distribution, the mean (\mean) and standard deviation (\sig),
to identify outliers. For a given threshold, $n$, the samples outside
$\mean\pm n\sig$ are treated as outliers. When the distribution is
affected by periodic RFI, measurements of \mean and \sig from the
data themselves are likely to deviate significantly from their true
values. Particularly, \sig is likely to be overestimated, which could
make the threshold based identification of outliers naturally less
efficient. For the example shown in Figure~\ref{fig-demo}, the
$\mean\pm3.5\sig$ bounds marked in the right panel for the original
data demonstrate how ineffective those would be in identifying any
outliers.
Using more robust methods, e.g., the iterative methods, computing
statistics using only RFI-free parts of the data, or using a reference
signal, \mean and \sig could be better estimated in some cases.
However, these better estimates would then imply identification
of significant fractions of data as outliers due to periodic RFI.
For the example shown in Figure~\ref{fig-demo}, more than 10\% of
the data would be rejected as outliers above $n=3.5$. Moreover, the
remaining fractions of data which are still contaminated by periodic
RFI but not identified as outliers, would leave their imprints in any
further analysis of such data.
\par 
The Fourier transform decomposes the input function into a sum of
complex exponentials
(i.e., complex sinusoids) of specific frequencies.
The amplitudes of the exponentials or the sinusoids would represent
the amount of power in the input function at the corresponding
frequency. For a discretely, uniformly sampled function $f(\rm t)$
of \nt samples, such as the time-sequence shown in Figure~\ref{fig-demo},
the Discrete Fourier Transform (DFT) may be written as
\begin{equation}
F(\nu_t) = \sum_{k=0}^{\nt-1} f(k\Delta t)\, e^{-2\pi i (\nu_tk\Delta t)},
\label{eq-dft}
\end{equation}
where $\Delta t$ is the sampling interval and $\nu_t$ is the Fourier
frequency. For real-valued function, $F(\nu_t) = F(-\nu_t)^\ast$.
The inverse DFT, which would transform $F(\nu_t)$ back to $f(\rm t)$,
is of similar form but with a normalization constant, 1/$\nt$, and
positive sign of the exponent.
For a signal or RFI that modulates sinusoidally with a period $P_s$,
the entire modulation power will be concentrated at $\nu_t=1/P_s$ in
the DFT. If the modulation envelope differs from sinusoid, the modulation
power gets distributed between the fundamental $\nu_t=1/P_s$ and its
harmonics. When the modulation period varies with time, the modulation
powers gets distributed in more than one
Fourier frequency bins
around the fundamental and its harmonics in the DFT.
A similar effect is seen when the input function length is a
non-integer multiple of the modulation period \citep[e.g., see][]{Ransom02}
In any case, power of a periodically modulated signal is concentrated within
a very limited number of bins
in the DFT than in the time sequence\footnote{An exception is when the
modulating signal approaches the Dirac comb function.}. Hence, it is
much easier to identify periodic signals in the Fourier domain.
\par
For a time-sequence of Gaussian distributed white-noise, the real
and imaginary parts of the Fourier transform also have Gaussian distributions
with $0$ mean and same standard deviations as of the input sequence. Since
all the power of any periodic RFI in time-domain will be contained within
much fewer frequency bins in the Fourier domain, it is much easier to identify the
corresponding samples in the DFT. If these identified outliers are replaced
by the expected mean of $0$ in the real and imaginary parts of the DFT, an
inverse DFT then would provide time-sequence wherein the periodic RFI is
effectively mitigated. An example of this mitigation is shown in
Figure~\ref{fig-demo} where the periodic RFI apparent in the second half of
the input sequence is successfully mitigated from the output sequence.
Note that the Fourier-domain mitigation of periodic RFI does not lead to
any loss of data samples in the time-domain. Even in the Fourier domain,
for the above example, only 0.6\% of the Fourier frequency space
was blanked. A more readily evident effect is that the underlying
distribution is restored (see the right panel), making the time sequence
much more suitable for conventional RFI excision. For the same thresholds,
only 0.8\% of samples in the Fourier domain-cleaned time sequence are found
to be outliers, as opposed to more than 10\% in the input sequence. More
importantly, restoration of the underlying distribution helps to identify
actual impulsive RFI samples which could have otherwise been hindered by
the periodic RFI modulation in the input sequence.
%
%
\begin{figure*}
\centering
\includegraphics[width=0.6\textwidth,angle=-90]{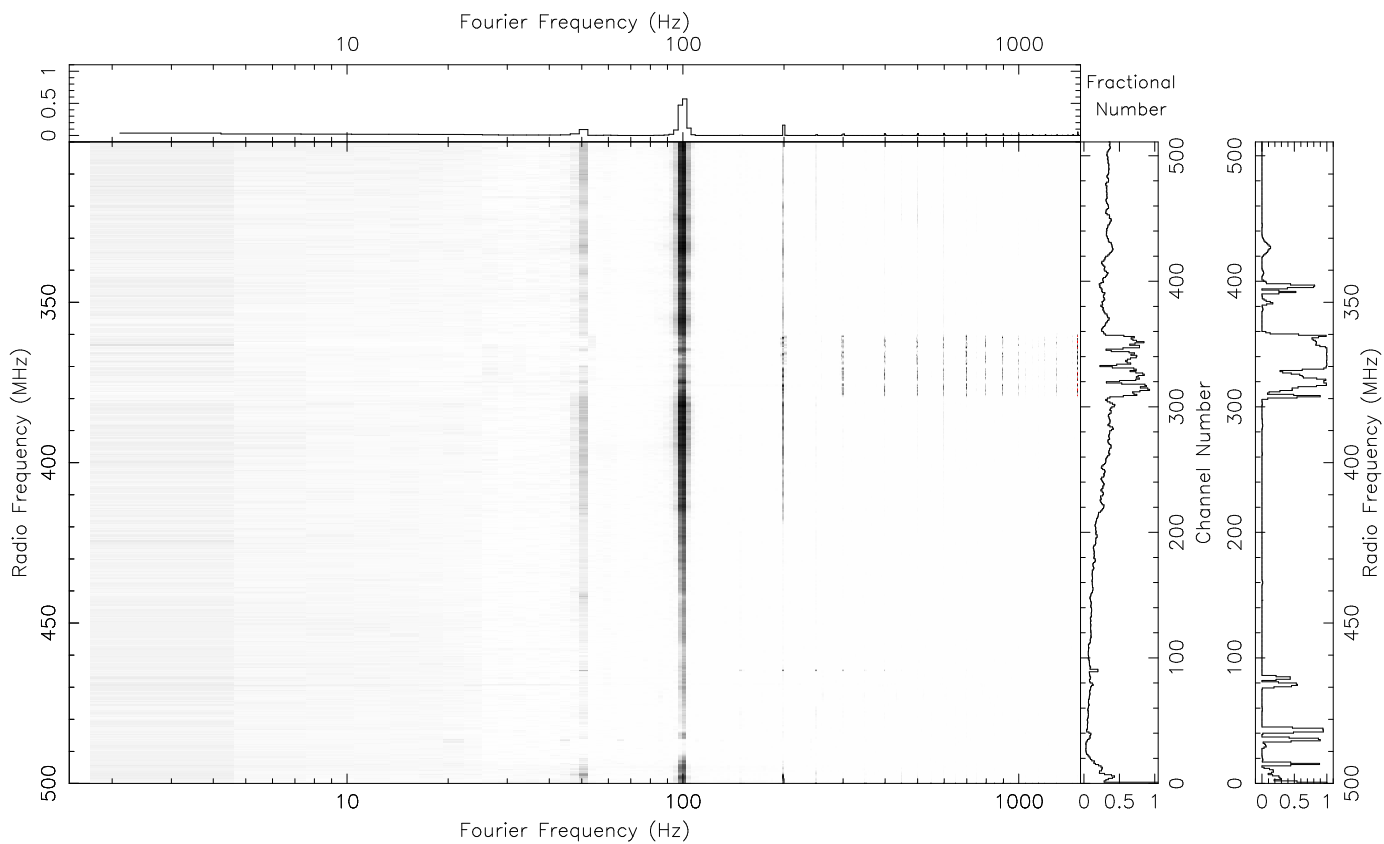}
\caption{A diagnostic plot output from \rficlean. The composite plot on the
left shows the statistics of the Fourier frequencies which were excised.
For every radio frequency channel in the main panel, the gray-scale mapping
shows how often a Fourier frequency was excised. Darker colors indicate
larger numbers. The adjacent top and right hand side panels show the averages
along the vertical and horizontal directions, respectively. The far right
hand side panel shows how often the radio frequency channels were
identified as narrow-band RFI and excised. See text for more details.}
\label{fig-diag}
\end{figure*}
%
\section{Mitigating periodic and other RFI from Time-domain data: \texttt{RFIClean}}
The above described method to mitigate periodic RFI has been
successfully used earlier in pulsar searches and studies
\citep[e.g.,][]{Faulkner04,SMK05,Camilo07z}. However, the usage has
generally remained limited to \emph{bandwidth-averaged} timeseries,
and for RFI with already known periodicities. As the strength,
and possibly even phase, of the periodic interference could be highly
variable across the observing bandwidth, its identification and mitigation
might not always be efficient in the bandwidth-averaged timeseries.
Moreover, mitigating some periodicities that are frequently caused by
known RFI (e.g., 50\,Hz) in the band-averaged timeseries
might also cause losing astronomical signals with unfavorably close by
fundamental or (sub-)harmonics
[e.g. the initial non-detection of PSR J2030+3641 by
\citet{Barr13} was potentially due to this fact].
Furthermore, with the ever growing number of RFI sources, it is not always
possible to know the RFI periodicity in advance.
With these considerations, we present below an approach to mitigate
periodic RFI (1) before averaging over the bandwidth, (2) only when
the RFI is strong enough to be detectable, and (3) without a priori
information on its periodicity.
\par
A widely used format for time-domain data recorded at radio
telescopes is the filterbank format, i.e., uniformly time-sampled streams
at a number of frequency channels across the observing bandwidth. In this
section, we describe a software package,
\rficlean\footnote{\url{https://github.com/ymaan4/rficlean}},
which we have developed to identify and mitigate periodic as well as
impulsive RFI from filterbank data.
The methods used are general purpose in the sense that no a priori
information about the periodicity or source of RFI is needed.
In the following subsections, we provide an operational overview of the
package followed by details on its computational and scientific performance
using real data.
\begin{figure}
\centering
\includegraphics[width=0.35\textwidth,angle=0]{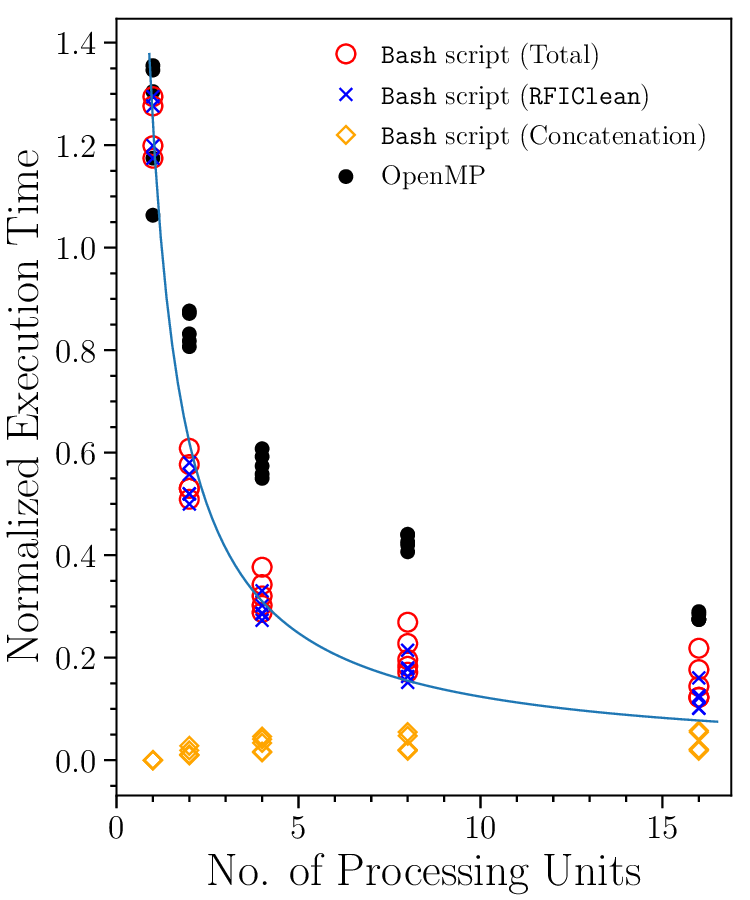}
\caption{Processing time spent per one second of observing duration as a
function of number of processing units are shown for two parallelization
methods. In the \texttt{bash}-script method, the time spent by the multiple
\rficlean instances, for concatenating the partial data products, and the
total time are shown separately using the red, blue and orange symbols.
The time spent using the OpenMP method is shown in black. The blue curve
shows the theoretical gain offered by an $n$-unit parallelization which
splits perfectly the workload, based on the average normalized execution
time obtained from the serial \texttt{bash}-script method (i.e., for a
single processing unit; 1.24 seconds/second).}
\label{fig-bench}
\end{figure}
%
\subsection{Operational overview and diagnostics}
\rficlean reads in and works on the input data block-by-block, where a
block represents time series of a user-specified length for a number of
radio frequency channels. Three categories
of RFI excision are employed. First in these categories is the Fourier
domain excision. For each time series, a DFT is computed using the
Fastest Fourier Transform in the West \citep[FFTW;][]{FFTW05} software
library.
The Fourier amplitude spectrum is often ``red'' (i.e., there is more
spectral power at lower frequencies than that at higher frequencies)
which prohibits efficient identification of any outliers. To avoid the
worsening effects of this redness, the amplitude spectrum is divided
in to several sections, whereas the length of each section is increased
logarithmically as we go from lower to higher Fourier frequencies. The
outliers are identified within each of these sections using robustly
computed mean and standard deviation, and a specified threshold. For each
of the outliers, the real and imaginary parts of the DFT are replaced
by 0s. An inverse DFT computed using FFTW then outputs a cleaned time
series. This operation is repeated for time series corresponding to
each of the radio frequency channels in the input block.
\par
Next, narrow band RFI is identified and excised from the above DFT-cleaned
block. For this, the mean and variance spectra are obtained for
the block, by computing robust mean and standard deviation of each of
the time series. The channels are identified which show either mean or
the variance power more than a specified threshold and termed as outliers.
The spectral power distribution in data obtained from radio telescopes
is often significantly affected by electronic filters and other parts
in the backends. This might make the above threshold based identification
of contaminated channels in the mean and variance spectra less effective.
To overcome any instrument induced spectral variations,
we compute the differences between the mean spectrum values at
adjacent channels, and denote the resultant spectrum as the
\textsl{difference-mean} spectrum. Hence, using the mean spectrum,
S$_{mean}$, the \textsl{difference-mean} spectrum, $\Delta$S$_{mean}$,
is computed as,
\begin{equation}
\Delta S_{mean}(j) = \sum_{j=0}^{\nf-2} [S_{mean}(j) - S_{mean}(j+1)],
\label{eq-diff}
\end{equation}
where \nf is the number of radio frequency channels. Note that
$\Delta$S$_{mean}$ consists of $\nf-1$ elements. With a
similar approach, we use the variance spectrum to compute the
\textsl{difference-variance} spectrum.
These difference-spectra, nearly free from the effects of the spectral
shapes induced by the instrument, are then used to further identify the
outlier channels. The time series in every identified outlier channel
is replaced by
the mean of the adjacent non-contaminated channel.
The spectra for the individual time samples are also examined to find
any narrow-band and narrow-time RFI, and the identified outliers are
replaced by the specified threshold.
\par
In the last excision step, the time series computed by averaging the
above DFT-and-spectral-cleaned block over all the channels is
then subjected to threshold based excision. This step identifies any
impulse-like RFI, and replaces the corresponding samples for each of
the frequency channels by the means of their corresponding time series.
\par
A diagnostic is output at the end, which primarily indicates the fractional
number of times a Fourier frequency was zapped for each of the radio
frequency channels, and how many times a channel was zapped in the
spectral cleaning. An example of a diagnostic plot using time-domain
data obtained from the giant metrewave radio telescope (GMRT) is shown in
Figure~\ref{fig-diag}. As evident from the main panel, periodic interference
at around 50\,Hz, and its first two harmonics (100 and 150\,Hz) were
zapped the most. The top panel suggests that the 100\,Hz interference
was identified and zapped from the time series for all the radio
frequency channels, on average, for about 50\% of the times. Corresponding
to this interference, the zapped frequencies are spread over 4$-$5 Fourier
bins, indicating the jitter in period of the interference. More than ten
harmonics of 50\,Hz are prominently visible particularly in the frequency
range 360$-$380\,MHz --- a part of the spectrum known to be often dominated
by interference at GMRT. As apparent from the far right panel in
Figure~\ref{fig-diag}, the spectral zapping block identified and excised this
fairly broad (about 20\,MHz) as well as several narrow band RFI.
\subsection{Computational performance}
Although \rficlean was primarily developed for excision of RFI in the offline
processing of data, the execution time could be of importance for processing
large data sets or if real-time excision of periodic RFI is desired. To
utilize more than one processing core, \rficlean offers two methods of parallel
processing. 

In the first method, a given filterbank data set 
is virtually divided into a number $n$ of time-segments, and individual serial instances 
of \rficlean are used to process these segments simultaneously and write the 
cleaned data in separate output files. At the end, the output files are 
concatenated in the correct order to output a single, RFI-excised filterbank 
file. A \texttt{bash} script is provided along with \rficlean which facilitates
the above parallel processing for a specified number of processing cores. 

For the second method, we first identified the most compute-intensive 
functions within \rficlean using the GNU
profiler\footnote{\url{https://sourceware.org/binutils/docs/gprof/}, 
last visited 11 November 2020.}. 
In the serial form of some of these functions, frequency channels are being processed independently from one another. Hence, it represents a case of a ``pleasingly parallel problem''. We modified data structures and parallelized sections of algorithms employing the OpenMP architecture\footnote{\url{https://www.openmp.org}, last visited 11 November 2020.}, such that channels can be processed and cleaned in parallel blocks
over $n$ threads.

We benchmarked the execution time using five filterbank
files parallelized over $n \in \{1,2,4,8,16\}$ processing units, where
a processing unit corresponds to the number of processes used in the
\texttt{bash}-script method, and the number of threads for the OpenMP
method, respectively. For this
benchmarking, we have used filterbank data with observing durations of about
5, 11, 22, 43 and 87 minutes, each with 2048 frequency channels and 
163.8\,$\upmu$s sampling time. \rficlean was specified to use 8192 samples
long time blocks to identify and excise RFI on a computing machine with
24 CPUs (2 Intel(R) Xeon(R) E5-2620v3 processors), 125\,GiB memory and
CentOS Linux 7 operating system.
We normalized the benchmarked execution times by the observing durations.

Results of the benchmark are shown in Figure~\ref{fig-bench}. For the first
method, the execution time scales down close to the expected linear
relationship (solid line) with the increasing number of processing cores.
While using a large number of cores, the speed-up in this method is primarily
limited by disk I/O.
The fractional time spent in concatenating the simultaneously processed data
into a single output also becomes significant while using large number of cores.
For the above mentioned sampling time and number of radio
frequency channels, \rficlean can excise periodic and other RFI in real-time
using just two processing cores. As the Fourier transform
computation time scales as $\Omega$(N log N), the execution time in
Figure~\ref{fig-bench} will also scale accordingly when a different block
size is used. Similarly for the second method, \rficlean operates in real
time starting as early as $n=2$. The execution time $T$ scales with the number
of threads, providing speed-ups ($T_{\mathrm{serial}}$/$T_{\mathrm{n}}$) up
to about 5 when using 16 threads. One can notice that this method is slightly
slower than the first method. This is primarily due to the fact that operations
related to input read and output write, as well as some of the not so
compute-intensive functions, are performed serially in the present version.
On the other hand, this method has the advantage of simplicity of execution
and setup while providing good acceleration.

\begin{figure}
\centering
\includegraphics[width=0.45\textwidth,angle=0]{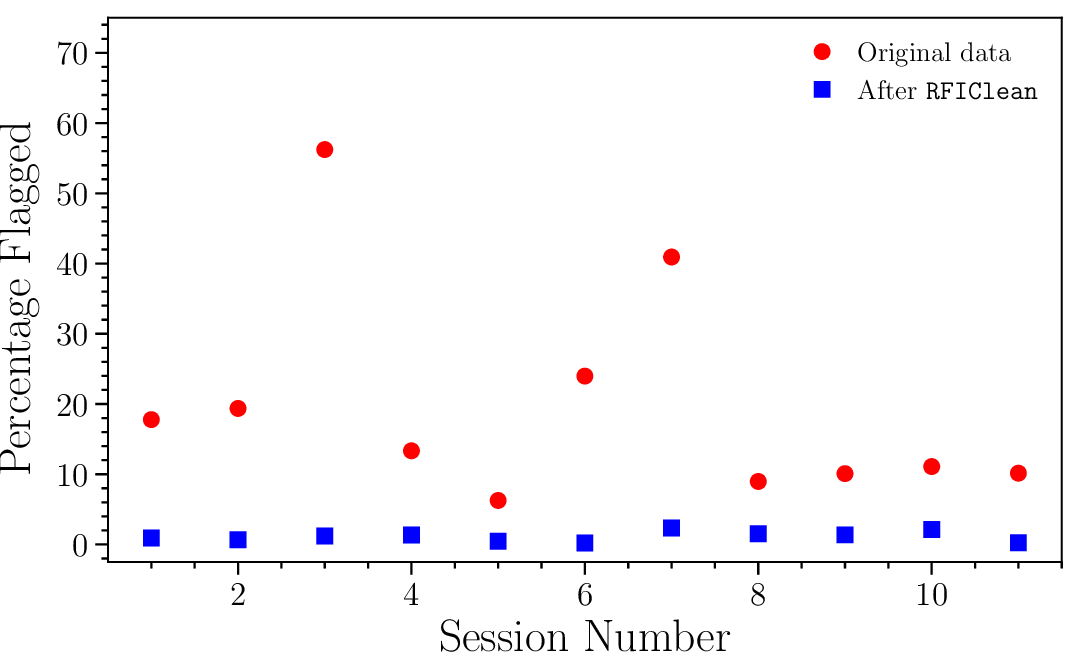}
\caption{The percentage of data that were masked by \texttt{rfifind} before
and after using \rficlean are shown for a number of observing sessions. A
stark improvement in the percentage of unmasked data is clearly evident.}
\label{fig-mask}
\end{figure}
\begin{figure}
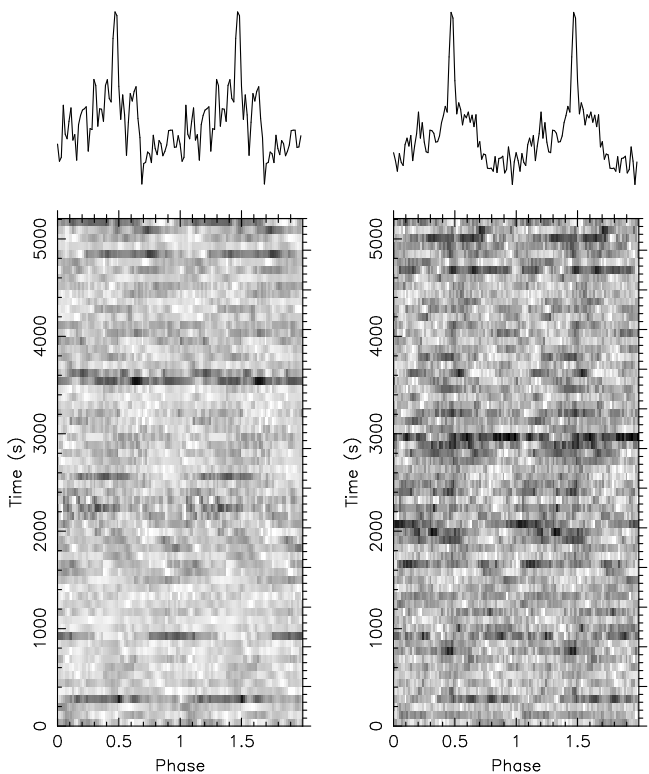

\centering
\subfigure{\includegraphics[width=0.55\textwidth,angle=-90]{he0532m4503_3_pstack_157.28ms_Cand.pfd.ps}}
\hspace*{2mm}
\subfigure{\includegraphics[width=0.55\textwidth,angle=-90]{he0532m4503_3_pstack_rficleaned_157.28ms_Cand.pfd.ps}}
 \caption{Detection of periodic signal from the pulsar J0533$-$4524
\citep{Oostrum20} before (left) and after (right) using \rficlean. 
The lower panels show a stack of partial average profiles obtained by
folding over the pulsar's period, with each row corresponding to about
80\,s of data, and the darker colors indicate higher intensities.
The upper panels display the vertical averages of data displayed in
the lower panels, and correspond to the net average profile of the
pulsar. Improvement in the detection significance of the pulsed
periodic emission is clearly visible.}
\label{fig-search}
\end{figure}
\begin{figure*}
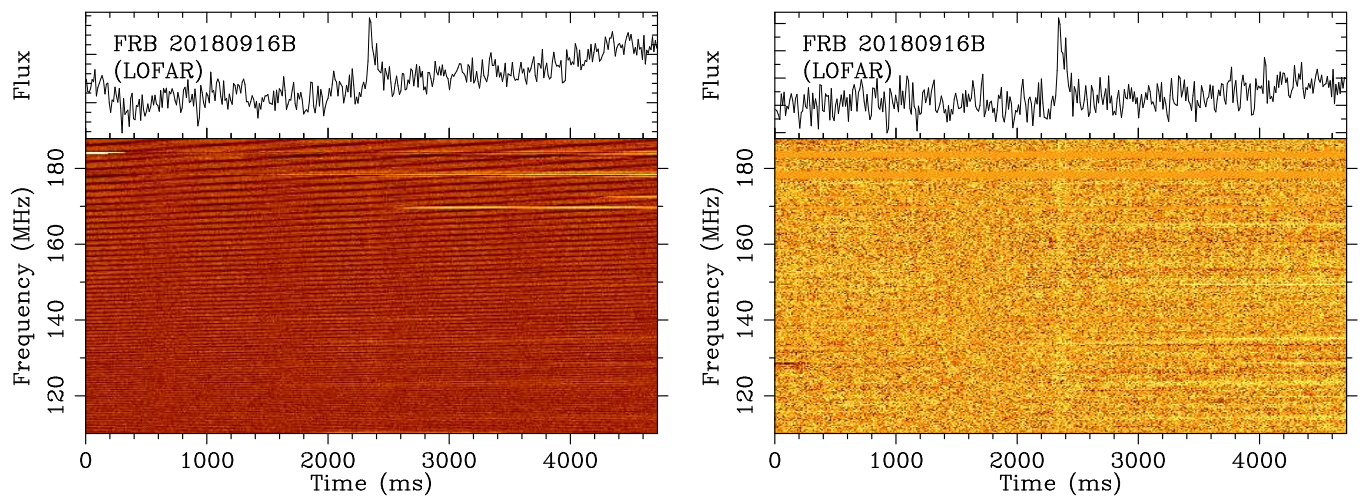

\centering
\subfigure{\includegraphics[width=0.35\textwidth,angle=-90]{l775795_deci_1620p3.ps}}
\hspace*{2mm}
\subfigure{\includegraphics[width=0.35\textwidth,angle=-90]{l775795_deci_deeprficlean_1620p3.ps}}
 \caption{Detection of a faint burst from the repeating \rfrb at
150\,MHz before (left) and after (right) using \rficlean.
The lower panels show the spectrograms around the burst arrival time,
after correcting for the known dispersive delay across the bandwidth.
Due to the dispersive delay corrections, the periodic RFI appears as
highly slanted stripes in the left image. 
The upper panels display the vertical averages of data displayed in
the lower panels, and correspond to the net average profile of the
burst. Improvement in the detection confidence and overall data quality
of the burst is clearly visible.}
\label{fig-frb}
\end{figure*}
\subsection{Scientific performance}
Here we demonstrate how \rficlean's usage could improve the fraction of
data that is lost in conventional RFI excision, and its efficacy in bringing
up faint astronomical signals (particularly in pulsar and FRB searches).
We also show that \rficlean can safeguard the periodic signals from
bright pulsars while mitigating periodic RFI, does not introduce any artifacts
on the pulse profiles obtained from cleaned data, and improves the quality
of the precision timing experiments.
\par
For some of the demonstrations below, we have used observations
of a very bright pulsar, \jpsr, conducted using GMRT on a number of epochs
between November 3, 2018 and March 1, 2019. In each observing session, data
were recorded with 2048 frequency channels and 0.16384\,ms sampling time,
and the observing durations varied between 2 and 6 minutes.
\subsubsection{RFI excision efficiency}
To demonstrate \rficlean's efficacy in improving the fraction of available
data that is not masked by conventional RFI excision methods, we have used
\texttt{rfifind} from \textsl{PRESTO}. \texttt{rfifind} examines data for
narrow and wide band interference
as well as strong periodic interference,
produces a mask for the parts of the data that it finds
RFI-contaminated and reports the masked percentage of the data.
For the above observations on \jpsr, the percentage of data that are masked
by \texttt{rfifind} before and after using \rficlean are shown in
Figure~\ref{fig-mask}. A sharp improvement in the percentage of the data
that \texttt{rfifind} finds RFI-free after mitigation using \rficlean is
clearly evident for all the observing sessions.
\subsubsection{Usefulness in pulsar and FRB searches}
In pulsar and FRB/fast-transient searches, the systematic noise due to RFI
reduces the S/N of the underlying astronomical signals as well as increases the
number of false candidates that a human or machine has to sift through
before identifying the genuine ones. The latter also decreases the
probability of any faint but genuine candidates to be identified correctly
with the required confidence. By efficient excision of periodic as well
as narrow/broadband RFI, \rficlean helps in uncovering the underlying
faint signals.
\par
Detection of the recently discovered pulsar J0533$-$4524
\citep{Oostrum20} at several epochs was in fact possible only after
mitigation of periodic RFI using \rficlean. An example of the improvement
in the pulsar's S/N after using \rficlean on a GMRT observation of this
pulsar in the 300$-$500\,MHz band is shown in Figure~\ref{fig-search},
where the formal S/N increased from less than 3 to about 25. Detection of
the magnetar J1809$-$1943 below 500\,MHz \citep{Maan19b} was also immensely
aided by \rficlean. In radio searches where information about the pulsar period is
already available \citep[e.g., from high-energy counterparts,][]{MA14,Maan15},
\rficlean can help in realizing high sensitivity by safeguarding the known
periodicity (see Section 3.3.3).
\begin{figure}
\centering
\subfigure{\includegraphics[width=0.48\textwidth,angle=0]{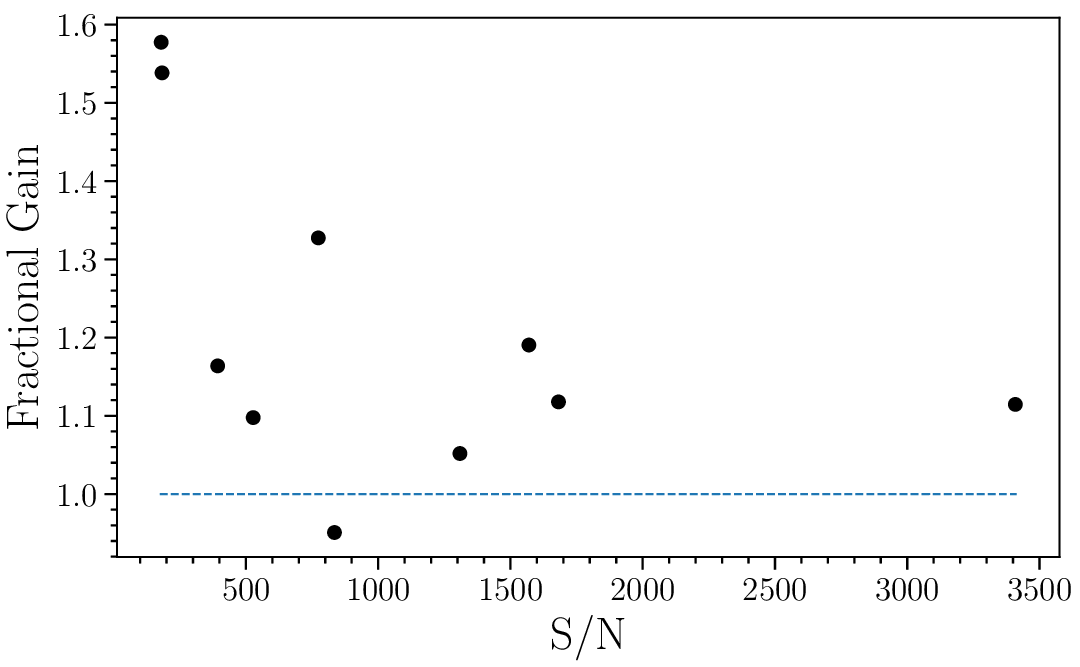}}
\subfigure{\includegraphics[width=0.48\textwidth,angle=0]{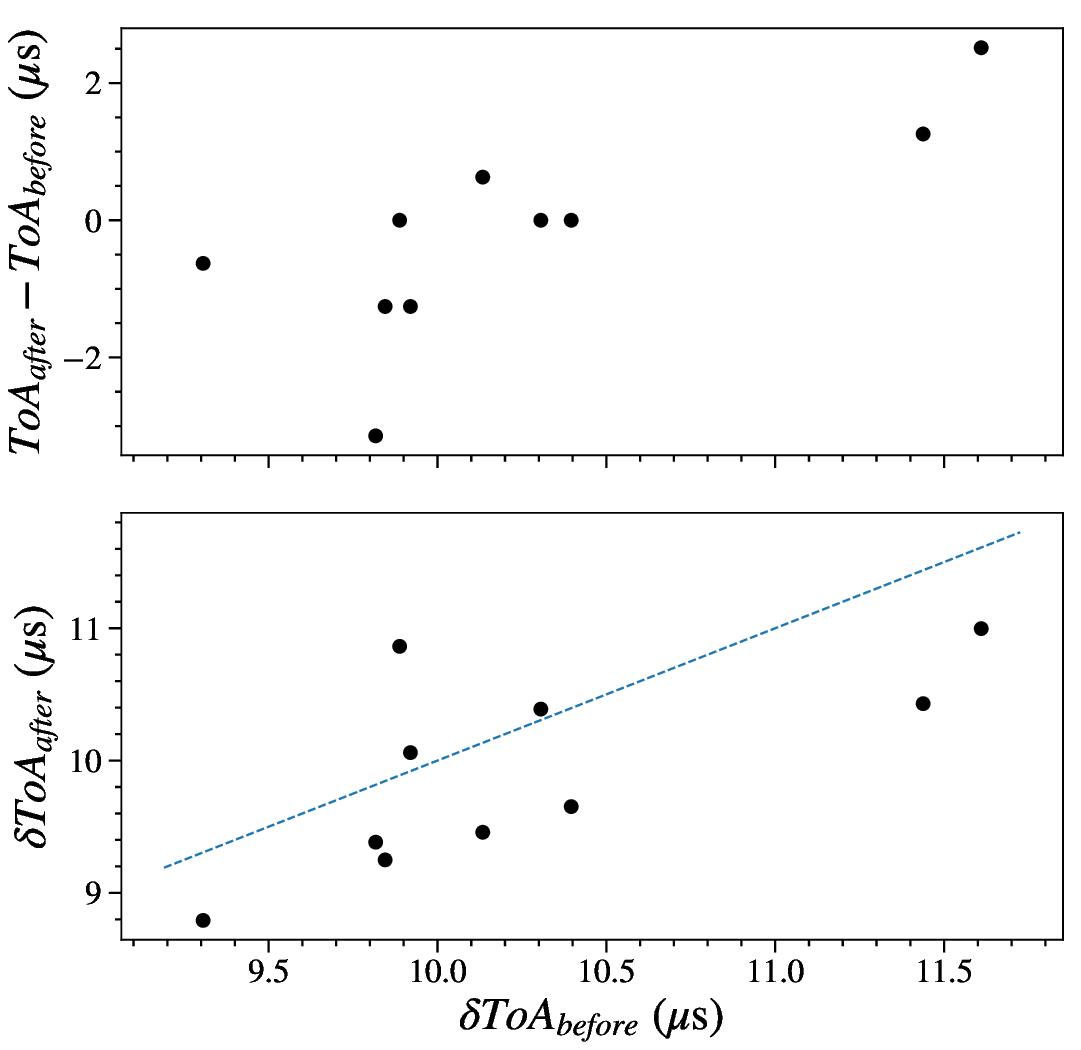}}
\caption{\textsl{\rficlean's suitability in timing studies:}
The upper panel shows the fractional increase in S/N after
excision using \rficlean, as a function of the S/N obtained from the
original data. The middle panel shows the difference in times-of-arrival
(ToA) measured from the original and the RFI-excised data, as a function
of the associated uncertainties in the original ToAs. The difference is
much smaller than the associated uncertainties. The bottom panel shows
a comparison between the ToA uncertainties measured before and after
the RFI excision. The uncertainties in measurements obtained from the
RFI-excised data are consistently smaller than their counterparts from
the original data for most of the observations.}
\label{fig-timing}
\end{figure}
\begin{figure*}
\centering
\subfigure{\includegraphics[width=0.75\textwidth,angle=0]{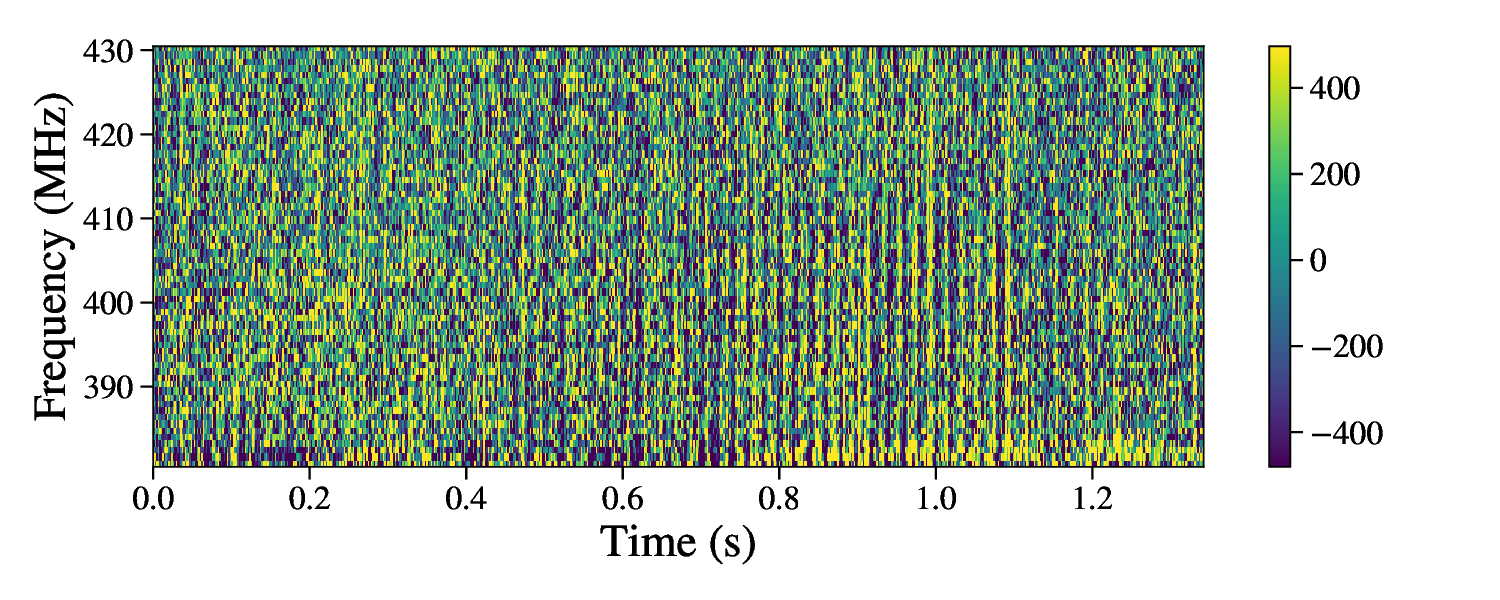}}
\subfigure{\includegraphics[width=0.75\textwidth,angle=0]{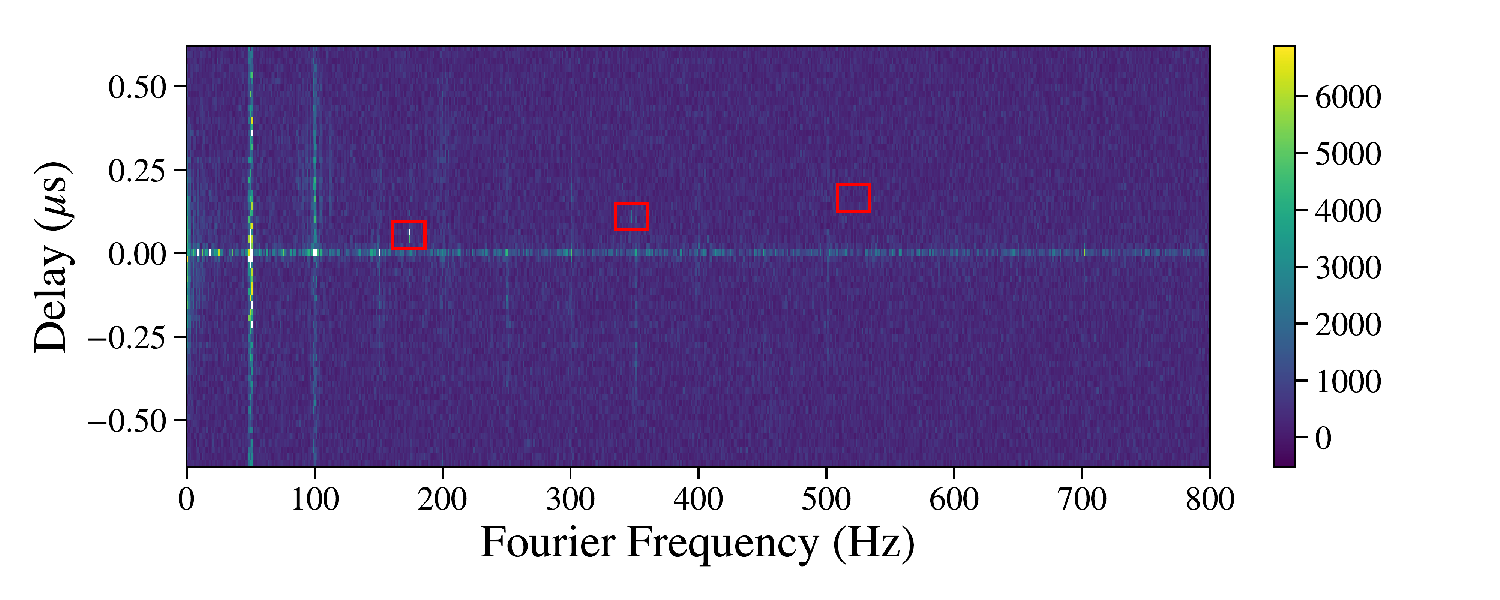}}
\subfigure{\includegraphics[width=0.75\textwidth,angle=0]{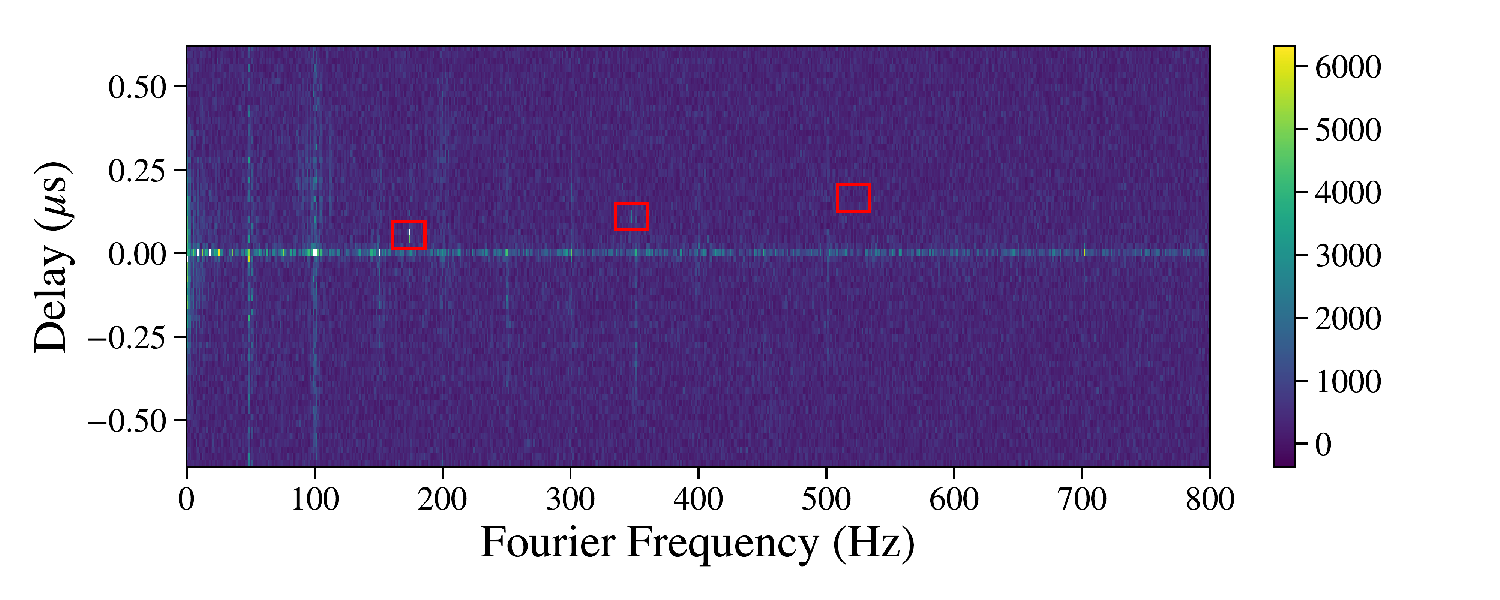}}
\subfigure{\includegraphics[width=0.75\textwidth,angle=0]{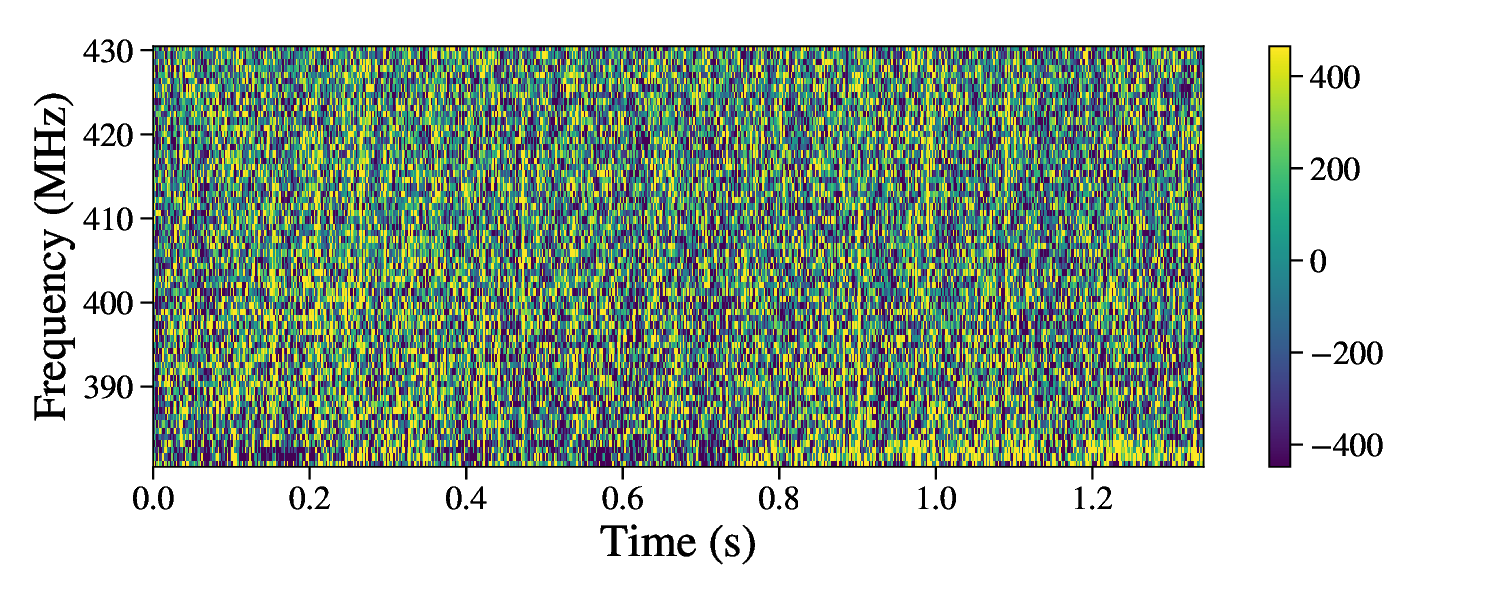}}
 \caption{The top panel shows spectrogram of a 1.3\,s of data from a GMRT
observation of the bright pulsar \jpsr, and with
visible contamination by periodic (50\,Hz) RFI. The second panel displays the
\tdft of the data in the top panel. Third panel displays the \tdft after the
data are already cleaned using \rficlean. The bottom panel shows the spectrogram
after the periodic RFI are identified and excised using the \tdft. The red-colored
rectangles in the middle panels mark the locations corresponding to the spin
period of \jpsr and its first two harmonics, after taking in to account the
vertical offset caused by the inter-channel dispersive delays.}
\label{fig-2dfft}
\end{figure*}
\par
More recently, a number of bursts from the repeating \rfrb have been
discovered at 150\,MHz using the Low Frequency Array (LOFAR) telescope
\citep{2020arXiv201208348P}, marking the first convincing detection
of any FRB at frequencies below 300\,MHz. The first of these detected
bursts (L01), which is also the faintest one, was confirmed only after some
strong periodic RFI were excised from the LOFAR data. A spectrogram
around the burst arrival time obtained from the original data as well as
that from data cleaned using \rficlean is shown in Figure~\ref{fig-frb}.
In the process of correcting for the large dispersive delay (nearly 79\,s!)
across the observed bandwidth for \rfrb, the originally undispersed periodic
RFI actually got dispersed and it appears in the form of highly slanted
stripes in the spectrogram. As evident in Figure~\ref{fig-frb}, \rficlean
successfully excised the periodic RFI
as well as some narrow-band RFI.
In this particular case, due to
very uniform amplitude of the periodic RFI across the bandwidth as well
as very large dispersive delay that had to be corrected, the RFI's effect
on the formal S/N of the burst was only marginal. However, confirmation
of the burst as an astronomical signal as well as study of its flux density
and spectrum were possible only after the RFI was mitigated using \rficlean.
The mitigation of periodic RFI from these data also reduced the false
burst candidates by nearly an order of magnitude.
\subsubsection{Usefulness in pulsar timing experiments}
Pulsars are excellent celestial clocks. Given the stability of their periods
up to one part in $10^{20}$, measuring the time-of-arrival (ToA) of their
pulses facilitate very high precision astrophysical
experiments such as testing the general relativity, constraining the
equation of state and detecting sub-$\upmu$Hz gravitational waves.
\par
Measuring the pulse ToA includes comparison with a template profile.
Any small change in the pulse-shape, intrinsic or that resulted from
the instrumental or processing software artifacts, would result in
an offset and/or increased uncertainty in the corresponding ToA
measurement. To avoid a possible
excision or modification of the periodic signal from strong pulsars,
\rficlean can safeguard the Fourier frequencies corresponding to a
pulsar's specified spin period.
\par
We demonstrate \rficlean's suitability
for high precision pulsar timing experiments using the observations
of the strong pulsar \jpsr discussed earlier in this section. As can
be seen in the top panel of Figure~\ref{fig-timing}, even without any
RFI excision the pulsar's strong periodic signal is detected with
high significance (S/N>200, after averaging over bandwidth and observing
duration) in all the observations. However, using
\rficlean further improves the detection significance for most of the
observations. The fractional improvement is naturally larger where the
original detection significance was smaller.
We also note that using \texttt{rfifind} on the data output
from \rficlean further improves the S/N by 10$-$20\% in many cases (not shown in
Figure~\ref{fig-timing}, but assessed separately).
\par
To generate a template, we used \texttt{psrsmooth} from \textsl{PSRCHIVE}
\citep{Hotan04}
on the profile with highest S/N. We then used \texttt{pat} on average
profiles obtained from all the observing sessions to compare with the
above template and measure the ToAs. The same template was used on the
original as well as RFI-excised data. As shown in the middle panel of
Figure~\ref{fig-timing}, the ToAs obtained from the original and
RFI-excised data are consistent with each other, and any difference
is much smaller than the associated uncertainty. This indicates that
\rficlean does not inflict any artifacts in the pulsar's profile
shape. The lower panel shows that, even for this strong pulsar, the
measurement uncertainties in the ToAs from the RFI-excised data are
smaller than their counterparts from the original data for most of
the sessions. The improvement in ToA uncertainty is a direct implication
of the enhancement in S/N shown in the upper panel. In fact much more
significant improvements in ToA uncertainties have been observed for
fainter pulsars, making \rficlean not only suitable but useful in
improving ToA measurements in pulsar timing experiments.
\par
\rficlean is an integral part of the data processing pipeline
\citep{Susobhanan20} used in the Indian Pulsar Timing Array (InPTA)
experiment. \rficlean is also being used in the pulsar timing
studies at the Argentine Institute of Radio astronomy \citep{2020arXiv201000010S}.
\section{Using two$-$dimensional DFT and other future work}
The identification and mitigation of periodic RFI in the previous section
is limited by the strength of such RFI within individual channels. If a
faint periodic RFI is spread over multiple channels or across the full
observed bandwidth, two-dimensional Discrete Fourier Transform (2D DFT)
might help in its identification and mitigation. While a detailed exploration
in this regard will be a subject of future work, here we present
a minimal demonstration of mitigating periodic RFI using \tdft.
\par
For a function discretely and uniformly sampled in two dimensions,
$f(\rm t, \rm f)$, such as the filterbank data discussed in the previous
section, the \tdft may be written as
\begin{equation}
F(\nu_t, \nu_f) = \sum_{k=0}^{\nt-1} \sum_{j=0}^{\nf-1} f(k\Delta t, j\Delta f)\, e^{-2\pi i (\nu_tk\Delta t+\nu_fj\Delta f)},
\label{eq-2dft}
\end{equation}
where $\Delta t$ and $\Delta f$ are the sampling intervals, \nt and \nf
are the total number of samples, and $\nu_t$ and $\nu_f$ are the corresponding
Fourier frequencies in the two dimensions. As for the one-dimensional DFT,
for a real-valued 2D function, $F(\nu_t,\nu_f) = F(-\nu_t,-\nu_f)^\ast$.
The inverse DFT, which would transform $F(\nu_t,\nu_f)$ back to $f(\rm t,\rm f)$,
differs in the signs of the two exponents
and a normalization constant, 1/($\nt\times\nf$).
\par
For the demonstration, we have used about 1.3\,s of data from one of
the GMRT observations on the bright pulsar \jpsr presented in the previous
section. To make the interference visually apparent in the spectrograms, the
already present interference by 50\,Hz (and its harmonics) were manually
enhanced by a factor of 4. The resultant spectrogram is shown in the top
panel in Figure~\ref{fig-2dfft}, and the second panel shows its \tdft.
Here $\nu_t$ is denoted as Fourier frequency and
$\nu_f\equiv\mathrm{Delay}$.
Interference at 50\,Hz and several of its harmonics is clearly visible.
Note that the variation in the interference power along the vertical
axis in this \tdft depends on how the interference power is distributed
along various frequency channels in the spectrogram. In an ideal case,
when the interference power is uniformly distributed across all the channels,
it will show up only on the $\mathrm{Delay}$=0 row in the \tdft.
\par
Note that the periodic signal from the pulsar \jpsr is dispersed due to
propagation effects in the ionized interstellar medium. As a consequence,
the periodic train of pulses originated from the source exhibits a
frequency dependent time delay. Due to this dispersed nature, the modulation
power of periodic signal from the pulsar is expected to be offset from the
$\mathrm{Delay}$=0 row in the \tdft. 
An offset from the $\mathrm{Delay}$=0 can also be used
to distinguish between RFI and astronomical dispersed signals, which
enables an efficient way of searching for new pulsars using \tdft.
\citet{Camilo96} used this method to search for millisecond
pulsars, and more details of the method can be found therein.
The expected locations corresponding to the pulsar's spin frequency and
first two harmonics are marked in the \tdft by red-colored rectangles in
Figure~\ref{fig-2dfft}. Even within this just 1.3\,s
of data, the excess power due to the pulsar signal is clearly visible
at the locations of the fundamental and first harmonic.
\par
The third panel of Figure~\ref{fig-2dfft} shows the \tdft after the
spectrogram was cleaned using \rficlean. While a noticeable decrease in
the interference power, primarily at 50\,Hz, is visible, it is evident
that the RFI is not fully mitigated. To mitigate the remnant RFI, we
computed the vertical average of the \tdft and identified the Fourier
frequencies where the power exceeded a specified threshold. The data
corresponding to these Fourier frequencies were replaced by local mean
values in the \tdft. The cleaned spectrogram obtained by an inverse
transform of the modified \tdft is shown in the bottom panel. While
this simple approach efficiently mitigated most of the periodic RFI
in this simple example, more complex contamination might require more
sophisticated approaches.
\par
New features for \rficlean are currently being planned,
including
reading and writing data in formats other than \textsc{sigproc}
filterbank, and implementing the spectral kurtosis method
\citep{NG10a}, to mitigate non-periodic RFI even more
efficiently.
That said, the current version has already shown to be a 
powerful and sometimes essential tool in discovering
and studying new pulsars and FRBs.

\section{Conclusions}
We have presented a method to identify and mitigate periodic RFI in the
Fourier-domain, and showed that mitigating periodic RFI is essential for
efficient identification and excision of other kinds of RFI. An accelerated
software implementation of the method, \rficlean, is publicly available at
\url{http://github.com/ymaan4/rficlean}. The method is shown to significantly
enhance the effective sensitivity, and we have presented a few examples where
it has been instrumental in detecting faint pulsars and FRBs. Given its
suitability and usefulness, the method is also already being used in pulsar
timing experiments. While our focus has been to show the benefits
on pulsar and FRB data, the method is directly applicable to a wide range
of radio observations, particularly in the time-domain, of diverse astrophysical sources.

\begin{acknowledgements}
We thank our referee, Scott Ransom, for his comments and suggestions
which have helped in improving the paper.
The research leading to these results has received funding from the European
Research Council under the European Union's Seventh Framework Programme
(FP/2007-2013) / ERC Grant Agreement n. 617199 (`ALERT'); from Vici research
programme `ARGO' with project number 639.043.815, financed by the Netherlands
Organisation for Scientific Research (NWO); and from the Netherlands eScience
Center (NLeSC) grant ASDI.15.406 (`AA-ALERT'). \\

This paper is based (in part) on data obtained with the International LOFAR
Telescope (ILT) under under project code COM\_ALERT, OBSID L775795.
LOFAR (van Haarlem et al. 2013) is the Low Frequency Array designed and
constructed by ASTRON. It has observing, data processing, and data storage
facilities in several countries, that are owned by various parties (each with
their own funding sources), and that are collectively operated by the ILT
foundation under a joint scientific policy. The ILT resources have benefitted
from the following recent major funding sources: CNRS-INSU, Observatoire de
Paris and Universit\'e d'Orl\'eans, France; BMBF, MIWF-NRW, MPG, Germany; Science
Foundation Ireland (SFI), Department of Business, Enterprise and Innovation
(DBEI), Ireland; NWO, The Netherlands; The Science and Technology Facilities
Council, UK; Ministry of Science and Higher Education, Poland.
\end{acknowledgements}

\bibliographystyle{aa}
\bibliography{references,rficlean_refs}


\label{lastpage}
\end{document}